\documentclass[showpacs,floats,floatfix,superscriptaddress,aps,prd,nofootinbib]{revtex4}
\usepackage{amsfonts}
\usepackage{amssymb}
\usepackage{amsmath}
\usepackage{bm}
\usepackage{xcolor}
\usepackage{epsfig}
\usepackage{ifthen}
\usepackage{graphicx}
\usepackage{multirow}
\usepackage{float}

\def\be{\begin{equation}}
\def\ee{\end{equation}}
\def\bea{\begin{eqnarray}}
\def\eea{\end{eqnarray}}
\def\bse{\begin{subequations}}
\def\ese{\end{subequations}}
\def\ba{\begin{array}}
\def\ea{\end{array}}

\def\i{\text{i}}

\def\ee{ \end{equation} }

\def\to{\rightarrow}

\def\a               {\alpha}
\def\b               {\beta}
\def\m             {\mu}
\def\n              {\nu}

\def\g              {\gamma}

\def\t {\tau}

\newcommand{\nn}{\nonumber\\}

\definecolor{darkgreen}{rgb}{0,.5,0}

\begin{document}

\author{I. Boradjiev}
\affiliation{Institute of Solid State Physics, Bulgarian Academy of Sciences, Tzarigradsko chauss\'{e}e 72, 1784 Sofia, Bulgaria}
\author{E. Christova}
\affiliation{Institute for Nuclear Research and Nuclear Energy, Bulgarian Academy of Sciences, Tzarigradsko chauss\'{e}e 72, 1784 Sofia, Bulgaria}
\author{H. Eberl}
\affiliation{Institute for High Energy Physics, Austrian Academy of Sciences,Vienna, Austria}

\email {boradjiev@issp.bas.bg, echristo@inrne.bas.bg,helmut.eberl@oeaw.ac.at}

\title{Dispersion theoretic calculation of the \boldmath $H\rightarrow Z+\gamma$ amplitude}

\begin{abstract}
We have calculated the $W$-loop contribution to the amplitude of the
decay $H\rightarrow Z+\gamma$ in two different methods: 1) in the
$R_\xi$-gauge using dimensional regularization (DimReg), and 2) in
the unitary gauge through the dispersion method. Using the
dispersion method we have followed two approaches: i) without
subtraction and ii) with subtraction, the subtraction constant being
determined adopting the Goldstone boson equivalence theorem (GBET)
at  the limit  $M_W\to 0$. The results of the calculations in
$R_\xi$-gauge with DimReg and the dispersion method with the GBET
completely coincide, which shows that DimReg is compatible with the
dispersion method obeying the GBET.
\end{abstract}

\pacs{14.80.BN, 
11.55.FV 
 }

\date{\today}


\maketitle

\section{Introduction}

The calculation of the Higgs decay rate into two photons through the
$W$-loop has become the subject of a controversy.
After extracting the transverse factor
\begin{equation} \label{Ptransverse}
{\cal P}_{\mu\nu}=k_{2\mu}k_{1\nu}-(k_1\cdot k_2) \, g_{\mu\nu},
\end{equation}
which takes current conservation into account, the invariant
amplitude is finite. Since however this amplitude is the sum of
individually divergent Feynman diagrams most authors use dimensional
regularization (DimReg) for its evaluation. Surprisingly, the DimReg
result \cite{EGN,SVVZ}\footnote{These are two among many such
calculations. The authors of \cite{MV} list 13 papers to which one
may add still another one, \cite{HS}, that also concurs with the
majority result.} differs (by a real additive constant) from the
outcome of a direct computation that works with the physical unitary
gauge \cite{GWW, GWW15}.

Responding to a criticism, \cite{W}, which points out that the
delicate cancellation of divergences is ambiguous and thus one needs
a regularization,   the result of \cite{GWW} was confirmed
in~\cite{CT}   by applying unsubtracted dispersion relations in a
calculation that  deals only with absolutely convergent integrals.
Nevertheless, this calculation was also subsequently criticized in
\cite{MV}. The origin of the controversy stems from the fact that
perturbative amplitudes may be ambiguous even if the corresponding
momentum space integrals are convergent: the Feynman rules need to
be supplemented by conditions like gauge invariance, or the
associated Ward identities,  alongside with locality (or causality
\cite{NST}) which yields the analytic properties in momentum space.
The argument for an unsubtracted dispersion relation follows directly from the fact that
the only constants that may appear in
perturbative calculations should be the coupling constants and masses that are
part of the full renormalizable Lagrangian.
Thus, the absence of  $H\g\g$-coupling in the SM Lagrangian,
implies a zero subtraction in the dispersion integrals
for the $H\rightarrow\gamma + \gamma$  amplitude.
The same argument holds for
the $H\rightarrow Z + \gamma$  amplitude, as well.

However,  since the SM is a spontaneously broken theory and
masses are generated through the Higgs mechanism, it was argued that
the considered amplitude  should obey the boundary condition defined by the Goldstone boson
equivalence theorem (GBET) \cite{GBET,J}.
In \cite{dedes2012} the amplitude of $H \to \gamma + \gamma$ was
calculated in the unitary gauge staying strictly in four dimensions but
fulfilling the Goldstone boson equivalence theorem.
Their result is the same as in \cite{EGN}.
In \cite{Wu_2017} it was
shown how the amplitude for the decay $H\rightarrow \gamma + \gamma$, calculated
in the $R_\xi$ gauge and in the unitary gauge, may lead to different results.

These controversial results in the calculations of the amplitude for $H\to \g +\g$ motivated us to consider
the decay $H\to Z +\g$. These two processes are similar  in a sense that
at tree level they are both zero and  induced by loop corrections only,
the $W$-loops giving the main contribution. At $M_Z=0$ the process $H\rightarrow Z + \gamma$ should reproduce the results for  $H\rightarrow \gamma + \gamma$.

In this paper we calculate the one-loop $W$-contributions
to $H\rightarrow Z + \gamma$ in two approaches: First we calculate the  amplitude using the dispersion relation approach.
We consider  two cases: 1) we assume the unsubtracted dispersion relation, and
2) we assume a non-zero subtraction constant adopting the GBET in the limit $M_W\to 0$.
Next we calculate the same amplitude in  the commonly used $R_\xi$-gauge using the conventional
dimensional regularization (DimReg).

The goal of these calculations is to compare the two results:
from the dispersion-relation approach, in which we deal with finite quantities only
- with  and without subtraction, to the result in  $R_\xi$-gauge with DimReg.
The dispersion-relation approach can, in fact, be viewed as a general tool
for resolving the ambiguities in the regularization scheme in
quantum field theory. We  show  that with the dispersion-relation
approach, where no regularization is necessary, and with subtraction
determined by the GBET,  we get
exactly the  same result as in $R_\xi$-gauge with DimReg.

Previously the decay $H\rightarrow Z + \gamma$ was calculated using
DimReg and $R_\xi$-gauge by Cahn \textit{et al.} \cite{CCF},
and later a complete analytic expression was obtained by other
authors \cite{BH,GKW}.
Recently in \cite{Hue_etal17} this calculation was done in the unitary gauge,
with the help of dimensional regularization.
We completely agree with their results.

Before we go into the details it shall be mentioned that in this study we have used a couple of helpful Mathematica packages,
\cite{Hahn_01,Hahn_99,Mert_91,Sh_16,Sh_17,Pat_15}.

\section{The Feynman diagrams} \label{section_FeynDiags}

\begin{figure*}[htb]\centering
\includegraphics[width=0.9\columnwidth]{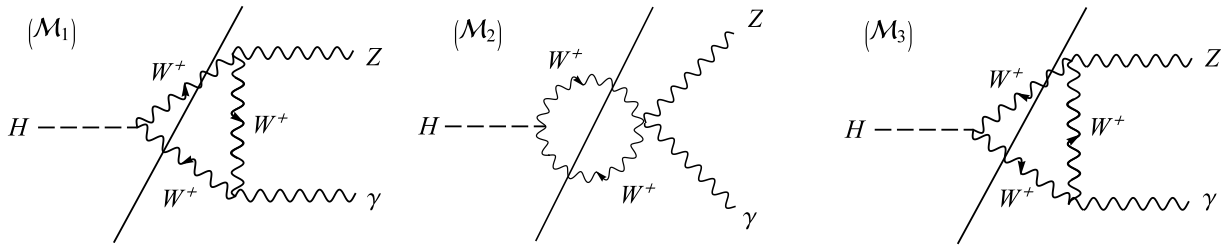}
\caption{Feynman diagrams for the $W$-loop contribution to the decay
$H\rightarrow Z + \gamma$. The inclined lines indicate the cuts.}
\label{Feynman_Diagrams}
\end{figure*}

We consider the contribution of the $W$-bosons loop-induced
amplitude of the decay $H\rightarrow Z + \gamma$. We work in the unitary
gauge, when only the physical particles contribute. There are
two types of diagrams. In Fig. \ref{Feynman_Diagrams}
the three $W$-loop diagrams that contribute to the absorptive part
of the amplitude are shown. These are the same diagrams as in the process
$H\to \gamma +\gamma$~\cite{GWW,CT}, in which  one of the final
photons is replaced by $Z$. In the same figure also the unitary
cuts, needed for obtaining the absorptive parts of the amplitude are
shown.
\\

\begin{figure*}[htb]\centering
\includegraphics[width=0.7\columnwidth]{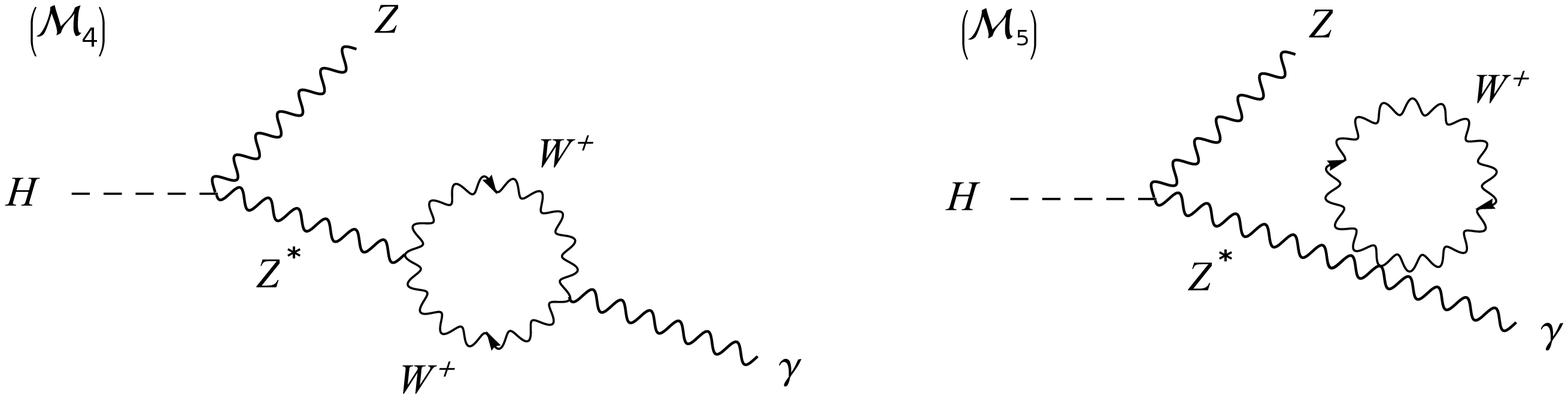}
\caption{ Feynman diagrams with an intermediate $Z^*$-boson for the
decay $H\rightarrow Z + \gamma$. Their contribution to the
absorptive part of the amplitude is  kinematically forbidden.}
\label{Feynman_Diagram_No_Contribution}
\end{figure*}

In Fig. \ref{Feynman_Diagram_No_Contribution} the two
additional diagrams that contribute to $H\to Z + \g$ are shown. These are
$H\to Z+Z^*$ with the subsequent transition $Z^* \to \gamma$ with
$W^+ W^-$ and $W^+$ in the loops. Clearly, kinematically their
contribution to the absorptive part is zero and we don't consider
them further.\\

\noindent
The amplitude for the process $\cal{M}$ is:
\begin{align}
{\cal M}=\mathcal{M}_{\mu\nu}(k_1,k_2)\,\zeta^{\mu}_1\,\zeta^{\nu}_2\, ,
\end{align}
where $k_1$ and $k_2$ are the  momenta of the $Z$-boson and the photon,
$\zeta_1, \, \zeta_2$ are their polarizations, orthogonal to $k_1$ and $k_2$, respectively:
\begin{align}
k_1^2 &= M_Z^2,
\qquad k_2^2 = 0,\qquad k_{1\mu}\zeta^{\mu}_1 = 0, \quad k_{2\nu}\zeta^{\nu}_2 = 0
\end{align}
The contribution to $\mathcal{M}_{\mu\nu} $ of the three diagrams
on Fig. \ref{Feynman_Diagrams} is:
\begin{align} \label{MmunuW}
\mathcal{M}_{\mu\nu} =
\mathcal{M}_{1\mu\nu} + \mathcal{M}_{2\mu\nu} + \mathcal{M}_{3\mu\nu}\, , \quad {\rm with}
\end{align}

\begin{widetext}
\begin{align}
\mathcal{M}_{1\mu\nu} &= \frac{-\i eg^2\cos\theta_WM}{(2\pi)^4}\int \text{d}^4k \,
\frac{V_{\mu\rho\beta}(-k_1, -P_2, P_1) \,
V_{\nu\gamma\sigma}(-k_2,-P_3, P_2)}{D_1 D_2 D_3} \times \notag \\
&\times \left(g_{\alpha}^{\beta} - \frac{P_{1\alpha}P_1^{\beta}}{M^2}\right)
\left(g^{\rho\sigma} - \frac{P_2^{\rho}P_2^{\sigma}}{M^2}\right)
\left(g^{\alpha\gamma} - \frac{P_3^{\alpha}P_3^{\gamma}}{M^2}\right),
\label{def_M1}
\end{align}
\begin{align}
\mathcal{M}_{3\mu\nu} &=
\mathcal{M}_{1\mu\nu}(\mu \leftrightarrow \nu, k_1 \leftrightarrow k_2) =  \notag \\
&= \frac{-\i eg^2\cos\theta_WM}{(2\pi)^4}\int \text{d}^4k \,
\frac{V_{\nu\rho\beta}(-k_2, -\tilde{P}_2, P_1) \,
V_{\mu\gamma\sigma}(-k_1,-P_3, \tilde{P}_2)}{D_1 \tilde{D}_2 D_3} \times \notag \\
&\times \left(g_{\alpha}^{\beta} -
\frac{P_{1\alpha}P_1^{\beta}}{M^2}\right) \left(g^{\rho\sigma} -
\frac{\tilde{P}_2^{\rho}\tilde{P}_2^{\sigma}}{M^2}\right)
\left(g^{\alpha\gamma} - \frac{P_3^{\alpha}P_3^{\gamma}}{M^2}\right),
\label{def_M3}
\end{align}
\begin{align}
\mathcal{M}_{2\mu\nu} &= \frac{\i eg^2\cos\theta_WM}{(2\pi)^4}\int \text{d}^4k
\frac{V_{\gamma\beta\mu\nu}}{D_1 D_3}
\left(g_{\alpha}^{\beta} - \frac{P_{1\alpha}P_1^{\beta}}{M^2}\right)
\left(g^{\alpha\gamma} - \frac{P_3^{\alpha}P_3^{\gamma}}{M^2}\right).
\label{def_M2}
\end{align}
\end{widetext}
Here, $\theta_W$ is the Weinberg (weak mixing) angle and $M=M_W$ is the mass of the W-boson.

The $WW\gamma$ and $WWZ$ vertices are denoted by $V_{\alpha\beta\gamma}$, the $WWZ\gamma$
vertex is denoted by $V_{\alpha\beta \mu\nu}$, they are given in Appendix \ref{secAA},
where all Feynman rules in the unitary gauge are recalled.

\noindent
We have also used the following brief notations:
\begin{align}
P_1 &= k + \frac{p}{2}, \qquad P_2 = k - \frac{v}{2}, \qquad  P_3 = k - \frac{p}{2}, \\
D_i &= P_i^2 - M^2 + i\epsilon, \qquad (i = 1,2,3), \\
\tilde{P}_2 &= k + \frac{v}{2}, \qquad \tilde{D}_2 = \tilde{P}_2^2 - M^2 + i\epsilon\\
p&=k_1+k_2,\qquad v=k_1-k_2\cdot
\end{align}
Taking into account the transformation properties under the reflection
$k \rightarrow -k$ of the loop momentum,
\begin{align}
\tilde{P}_2 (k \to -k) = -P_2, \qquad \tilde{D}_2(k \to -k) = D_2 \, .
\end{align}
We relate $\mathcal{M}_{3\mu\nu}$ and $ \mathcal{M}_{1\mu\nu}$,
thus simplifying our calculation:
\begin{align}
\mathcal{M}_{3\mu\nu} (k \to -k) &= \mathcal{M}_{1\mu\nu} \, . \label{M3_reflection_symm}
\end{align}

\section{Absorptive part of the amplitude} \label{AbsorptivePartAmplitude}

\noindent
We obtain the absorptive part through the Cutkosky rules
which set the momenta of the $W$'s on-shell \cite{C}:
\begin{align}
\frac{1}{p^2 - M^2} \longrightarrow \,
(2\pi\i) \, \theta(\pm p_0) \, \delta(p^2 - M^2)\, .
\label{Cutkosky-trans}
\end{align}
The imaginary part is obtained via the cut diagrams, $\mathcal{M}_{i\mu\nu}^C$:
\begin{align}
\Im m \,\mathcal{M}_{\mu\nu} &= -\frac{\i}{2} \,
(2\mathcal{M}_{1\mu\nu}^C + \mathcal{M}_{2\mu\nu}^C)\, .\label{ImM}
\end{align}
Obviously, here we have taken into account Eq. (\ref{M3_reflection_symm}).

\noindent
Further we define the invariant absorptive  part ${\cal A}$ of the amplitude
through the imaginary part of the amplitude:
\begin{align}
\Im m \, \mathcal{M}_{\mu\nu} = \frac{eg^2\cos\theta_W}{8\pi M}
\, \mathcal{A}(\tau) \mathcal{P}_{\mu\nu},
\qquad \tau = \frac{p^2}{4M^2},
\end{align}
where $\mathcal{P}_{\mu\nu}$ is the transverse-momentum (gauge invariant), given by Eq.~(\ref{Ptransverse})
,
\begin{align}
& k_1^{\mu} \mathcal{P}_{\mu\nu} =  k_2^{\nu} \mathcal{P}_{\mu\nu} =0.
\end{align}
Then ${\cal A}$ is obtained via the expression:
\begin{align}
\mathcal{A}(\tau) \mathcal{P}_{\mu\nu}
&= \frac{M^2}{\pi} \int \text{d}^4k \,
\mathcal{I}_{\mu\nu} \, \theta(P_{10})\theta(-P_{30})\delta(D_1)\delta(D_3),
\label{Def_Absorptive_Amplitude}
\end{align}
where $\mathcal{I}_{\mu\nu}$ is determined by the Feynman diagrams on Fig. \ref{Feynman_Diagrams}.
The two delta functions $\delta(D_1)$ and $\delta(D_3)$ in Eq.~(\ref{Def_Absorptive_Amplitude})
reduce the one-loop integral to a phase-space integral. In the next section as the second step we will calculate
from the absorptive part the real part of the amplitude by applying the dispersion integral technique.
One can also inverse the step of computing the absorptive part. Instead of cutting the one-loop amplitude,
one can sew appropriate tree-level amplitudes together to form the one-loop amplitude which turns the
cutting step around, avoiding the explicit construction of one-loop Feynman diagrams. But then one can rely on
evaluating Feynman integrals to do the second step \cite{new0}. These are the so-called unitarity cut methods based on \cite{new1},
see also e. g. \cite{new2, new3}.

\noindent
The tensor $\mathcal{I}_{\mu\nu}$ is obtained via straightforward, but rather tedious
calculations starting from the expressions (\ref{def_M1})-(\ref{def_M2}).
Also we make use of the following identities, that hold for both
the $WW\gamma$ and $WWZ$ vertices:
\begin{align}
V_{\alpha\beta\gamma}(p_1, p_2, p_3) &= - V_{\beta\alpha\gamma}(p_2, p_1, p_3) =
V_{\gamma\alpha\beta}(p_3, p_1, p_2),
\end{align}
and
\begin{align}
& p_1^{\alpha}V_{\alpha\mu\gamma}(p_1,-k_1,p_3) =
p_3^2g_{\mu\gamma} - p_{3\mu}p_{3\gamma} - M_Z^2g_{\mu\gamma} , \\
& p_1^{\alpha}V_{\alpha\nu\gamma}(p_1,-k_2,p_3) =
p_3^2g_{\nu\gamma} - p_{3\nu}p_{3\gamma}, \\
& p_1^{\alpha}p_3^{\gamma}V_{\alpha\mu\gamma}(p_1,-k_1,p_3) = -M_Z^2P_{3\mu}, \\
& p_1^{\alpha}p_3^{\gamma}V_{\alpha\nu\gamma}(p_1,-k_2,p_3) = 0.
\end{align}

\noindent
After rather cumbersome calculations we end up with the following expression
for $\mathcal{I}_{\mu\nu}$ :
\begin{widetext}
\begin{align}
\mathcal{I}_{\mu\nu}&=
\frac{8M_Z^2}{M^4D_2} \, k^2\left(k_{\mu}k_{\nu} + \frac{k_{2\mu}k_{\nu}}{2} -
\frac{k_{\mu}k_{1\nu}}{2} - \frac{k_{2\mu} k_{1\nu}}{4}\right) + \frac{-2M_Z^2}{M^4} \,
k^2g_{\mu\nu} \notag \\
&+ \frac{8M_Z^2}{M^2D_2} \, \left[-k_{\mu}k_{\nu} - \frac{k_{2\mu}k_{\nu}}{2} +
\frac{k_{\mu}k_{1\nu}}{2} - \frac{k_{2\mu} k_{1\nu}}{8} +
\frac{1}{4} \, g_{\mu\nu}k_1\cdot k_2 - \frac{1}{8} \, g_{\mu\nu}k\cdot (k_1-k_2)\right]   + \frac{M_Z^2}{M^2} \, g_{\mu\nu} \notag \\
&+ \frac{2}{M^2D_2} \left[4 k_1\cdot k_2 \, k_{\mu}k_{\nu} + 2k^2k_{2\mu}k_{1\nu}
- 4k\cdot k_1 \, k_{2\mu}k_{\nu} - 4 k\cdot k_2 \, k_{\mu}k_{1\nu}\right. \notag \\
&\left.\hspace*{7cm}+ g_{\mu\nu}\left(4k\cdot k_1 \, k\cdot k_2 - 2k^2 \,
k_1\cdot k_2\right) \right]  \notag \\
&+ \frac{2}{D_2} \left[ \left(-3k^2 + 3k\cdot k_1 - 3k\cdot k_2 -\frac{9}{2} \,
k_1\cdot k_2 + 3M^2 - \frac{3}{4} \, M_Z^2\right)g_{\mu\nu} \right. \notag \\
&\left.\hspace*{7cm} + 12k_{\mu}k_{\nu} + 3k_{1\nu}k_{2\mu} - 6k_{\mu}k_{1\nu} +
6k_{2\mu}k_{\nu} \right]. \label{Integrand_I}
\end{align}
\end{widetext}
Now we have to do the integration in (\ref{Def_Absorptive_Amplitude}).
We perform it in the rest frame of the decaying Higgs boson, with the $z$-axis
pointing along  $\textbf{k}_1$:
\begin{align}
p^{\alpha} &= k_1^{\alpha} + k_2^{\alpha} = (p, \mathbf{0}),\qquad p\equiv p_0 = 2M\sqrt{\tau}, \\
k_1^{\alpha} &= \frac{p}{2\tau} (\tau+a, 0, 0, \tau-a) , \quad
k_2^{\alpha} = \frac{p}{2\tau} (\tau-a, 0, 0, a-\tau) ,
\quad a = \frac{M_Z^2}{4M^2}=\frac{1}{4\cos^2\theta_W}  , \\
v^{\alpha} &= k_1^{\alpha} - k_2^{\alpha} = \frac{p}{\tau} (a, 0, 0, \tau-a) , \quad
v^2 = 4M^2 (2a-\tau), \quad (p\cdot v) = M_Z^2 .
\end{align}
The two $\delta$-functions: $\delta(D_1) = \delta[(k+p/2)^2-M^2]$ and
$\delta(D_3) = \delta[(k-p/2)^2-M^2] $  immediately determine $k_0$ and $\vert \mathbf{k}\vert$:
\begin{align}
& k^{\alpha} = (k_0,\text{k}) \Rightarrow k_0 = 0, \,
\vert \mathbf{k}\vert ^2 = M^2(\tau-1) = \frac{p^2}{4} \, \beta^2,
\end{align}
where
\begin{align}
\beta = \sqrt{1-\tau^{-1}}\, .
\end{align}
Thus, we are left only with the 2-dimensional integral over the direction of
$\mathbf{k}=\vert \mathbf{k}\vert (\sin\theta\cos\phi ,\sin\theta\sin\phi ) $.
For $D_2$ we obtain:
\begin{align}
 D_2 = -2M^2 (\tau-a) (1 - \beta\cos\theta) \, .
\end{align}

\noindent
The absorptive part of the amplitude is non-zero at $\tau > 1$ and it reads:
\begin{align}
&\mathcal{A}(\tau) = \,  \frac{a}{\tau-a} \left\{
\left[ 1 + \frac{1}{\tau-a} \left( \frac{3}{2} - 2a\tau \right) \right] \beta \, - \, \right. \notag \\
& \left. \left[1 - \frac{1}{2(\tau-a)} \left(2a-\frac{3}{2\tau}\right) - \frac{3}{2a}
\left(1-\frac{1}{2\tau}\right)\right] \ln \left(\frac{1+\beta}{1-\beta}\right) \right\} , \notag \\
& \qquad\qquad \tau > 1\,  .
\end{align}
The details of the calculations are presented in Appendix \ref{secAB}.

\section{Real part of the amplitude} \label{}

\noindent
The full invariant amplitude  $\mathcal{F}(\tau ,a)$ is defined  by
\begin{align}
\mathcal{M}_{\mu\nu} = - \frac{e g^2\cos\theta_W}{8\pi M} \,
\mathcal{F}(\tau ,a) \mathcal{P}_{\mu \nu},\label{ReM}
\end{align}
where $\mathcal{P}_{\mu \nu}$ is the transverse-momentum factor (\ref{Ptransverse}).

The vanishing of the absorptive part of the amplitude at $\tau <1$ tells us that
the invariant amplitude $\mathcal{F}$ at $\tau <1$, which is the physically interested region, is only real.
Following the analytic properties of the amplitude, we define the invariant {\it unsubtracted} amplitude ${\cal F}_{un}(\tau ,a)$ in this region,
 $\tau <1$, through  the convergent dispersion integral:
\begin{align}
\mathcal{F}_{un}(\tau ,a) = \frac{1}{\pi}
\int_{1}^{\infty}\frac{\mathcal{A}(y)}{y-\tau} \, \text{d}y, \qquad \tau <1. \label{DI}
\end{align}

From the explicit expression for $\cal A$ and its behaviour at $\tau \to \infty$
we obtain that this integral is absolutely convergent.
This however does not imply that there are no subtractions in (\ref{DI}):
the dispersion integral (\ref{DI}) determines the {\it full} amplitude ${\cal F} (\tau ,a)$ up to an additive  constant $C(a)$:
\be
2\pi\,\mathcal{F}(\tau ,a)= 2\pi\,\mathcal{F}_{un}(\tau ,a) +{\cal C}(a)\cdot\label{Ffull}
\ee

In order to determine ${\cal C}(a)$ we need some additional information about the amplitude --
some boundary condition or a physical measurable quantity at some fixed value of $\tau$. In our calculations
we fix ${\cal C}(a)$ through the Goldstone Boson equivalence theorem (GBET)~\cite{GBET}, which fixes the behaviour of the amplitude at $\t\to\infty$.\\

\noindent
In accordance with this we  calculate the amplitude ${\cal F}(\tau ,a)$ in two steps:

1. First we calculate ${\cal F}_{un}(\tau ,a)$ using the dispersion relation Eq. (\ref{DI}).

2. We calculate ${\cal C}(a)$ using  the GBET.

\subsection{The unsubtracted amplitude \boldmath ${\cal F}_{un}(\tau ,a)$}

The unsubtracted amplitude ${\cal F}_{un}(\tau ,a)$ is determined by the convergent dispersion integral Eq.~(\ref{DI}).
The integrals in Eq.~(\ref{DI}) are taken analytically - they are given in Appendix \ref{secAC}, and we obtain:
\begin{align}\label{amplitude_F}
 & 2\pi\mathcal{F}_{un}(\tau ,a) = \,\, \frac{3 - 4a^2}{\tau-a}  \, + \notag \\
 & \left( 6-4a - \frac{3 - 4a^2}{\tau-a} \right) F(\tau, a)
- 2a \left( 2 + \frac{3 - 4a\tau}{\tau-a} \right) G(\tau, a)\, ,
\end{align}
\begin{align}
&F(\tau,a) = \frac{f(\tau) - f(a)}{\tau - a} \, , \\
&G(\tau,a) = \frac{g(\tau) -  g(a)}{\tau-a} \, ,\\
\notag\\
f(x) = &\left\{
\begin{tabular}{r}
\quad $ \arcsin^{2}(\sqrt{x}) \qquad\qquad\qquad$ for $\, x \leq 1$, \\
\\
$ -\frac{1}{4}\left(\ln \frac{1 + \sqrt{1 - x^{-1}}}{1 - \sqrt{1 - x^{-1}}}
  - i\pi\right)^2 \qquad$ for $\, x > 1$, \\
\end{tabular}
       \right. \\
\notag \\
g(x) = &\left\{
\begin{tabular}{r}
$ \sqrt\frac{1-x}{x}\, \arcsin(\sqrt{x}) \quad\quad\qquad\qquad$ for $\, x \leq 1$, \\
\\
$ \frac{1}{2} \sqrt\frac{x-1}{x} \, \left(\ln \frac{1 + \sqrt{1 - x^{-1}}}{1 - \sqrt{1 - x^{-1}}}
- i\pi\right) \qquad$ for $\, x > 1$. \\
\end{tabular}
      \right.
\end{align}

The result for $\tau > 1$ in the above formula is obtained via analytic continuation.
(The same result may be found if we had set $\tau > 1$ in the integrand and taken
the $i\epsilon$ prescription in $D_2$ into account.)\\

\noindent
There are several important physical consequences for this amplitude.
\begin{enumerate}
\item  The amplitude at threshold, $\tau = a$, is finite. We have:
\begin{align}
&\lim _{\tau\rightarrow a}2\pi\mathcal{F}_{un}(\tau ,a) = \frac{1}{2a} \left[3 - 4a + 4a^2 -\frac{(3 - 16a +12a^2)}{1-a} \, g(a) \right].
\end{align}
The  absence of singularities in the amplitude is in accordance with the required analytic
properties of $\mathcal{F}_{un}(\tau ,a)$, necessary for the validity of the dispersion relations.\\

\item In the asymptotic limit $\tau \to\infty$, which implies $M^2_H \gg M^2$ at fixed $a$, we obtain:
\begin{align}\label{Finfty}
\lim_{\tau \to \infty} \, \mathcal{F}_{un} (\tau ,a) = 0.
\end{align}

\item In the limit of $a \rightarrow 0$, we have to recover the corresponding  invariant
amplitude $\mathcal{F}^{\gamma\gamma} (\tau)$ for the $H\rightarrow\gamma + \gamma$ process:
\begin{align}
\mathcal{M}^{\gamma\gamma}_{\mu\nu} =
\frac{-e^2 g}{8\pi M}\,\mathcal{P}_{\mu\nu}\,\mathcal{F}^{\gamma\gamma} (\tau),\label{calF}
\end{align}
where $\mathcal{P}_{\mu\nu}$ is the same transverse bilinear combination as
Eq. (\ref{Ptransverse}) with the (on shell) photon momenta $k_1, \, k_2$.
We obtain:
\begin{align}
\lim_{a\rightarrow 0} 2\pi \mathcal{F}_{un}(\tau ,a) =
3\tau^{-1}[1+(2-\tau^{-1}) \, f(\tau)],
\end{align}
which is exactly the result for $F^{\gamma\gamma}_{un}(\tau)$, obtained in the unitary gauge, both,
with direct calculations without renormalization in \cite{GWW},
and using the dispersive relations approach without subtraction in \cite{CT}.

\item  We calculated also the amplitude of the process in the commonly used $R_\xi$-gauge using DimReg.
The calculation was done with the help of the automatic
tools FeynArts~\cite{Hahn_01}
and FormCalc~\cite{Hahn_99}. There
are 20 Feynman triangle vertex graphs, 6 Feynman vertex graphs with a four-point interaction
and 10 graphs with selfenergies from $Z^*-\gamma$
transition. It is checked that the result is
UV finite and independent of $\xi$ and it coincides with the
one, obtained earlier in \cite{BH}.
However, the result for the amplitude
$\mathcal{F}_{\text{DimReg}}(\tau)$,  obtained using dimensional
regularization,  differs by a real  additive constant from our result for $\mathcal{F}_{un}(\tau)$:
\begin{align}
2\pi\mathcal{F}_{\text{DimReg}}(\tau ,a) =  2\pi\mathcal{F}_{un}(\tau ,a) + 2(1-2a),
\label{differs}
\end{align}
which leads to a non-vanishing asymptotic behaviour at $\tau\rightarrow \infty$.

\end{enumerate}

\begin{figure*}[htb]\centering
\includegraphics[width=0.9\columnwidth]{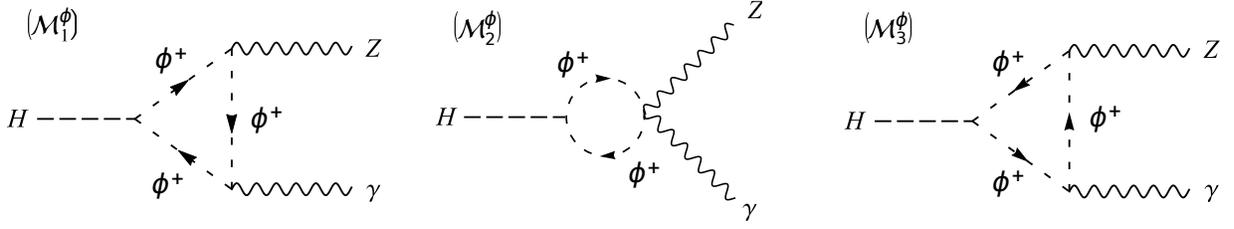}
\caption{ The vertex Feynman diagrams for the charged Higgs ghost contribution to the decay
$H\rightarrow Z + \gamma$.
The Cutkosky cuts are analagous to those shown in Fig.~\ref{Feynman_Diagrams}.
\label{Feynman_Diagrams_phi}
}
\end{figure*}

\begin{figure*}[htb]\centering
\includegraphics[width=0.7\columnwidth]{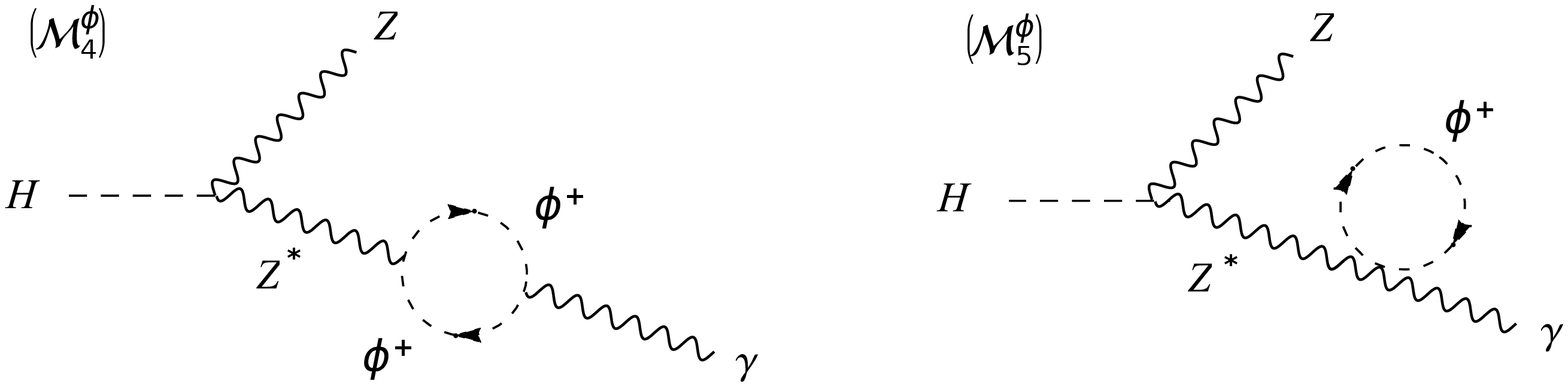}
\caption{ The selfenergy Feynman diagrams  for the charged Higgs ghost contribution with an intermediate $Z$-boson (in the unitary gauge) for the decay $H\rightarrow Z + \gamma$.
Their contribution to the absorptive part of the amplitude is zero, being kinematically forbidden. \label{Feynman_Diagram_No_Contribution_phi}
}
\end{figure*}

\subsection{The charged ghost contribution and the  constant \boldmath ${\cal C}(a)$}

We determine the  constant ${\cal C}(a)$ through the charged ghost
contribution adopting the GBET, which implies that at $M_W \to 0 $,
i.e. at $\tau \to \infty$,
 the $SU(2)\times U(1)$ symmetry of the SM is restored
and the longitudinal components of the physical $W^\pm$-bosons
 are replaced by the physical Goldstone bosons $\phi^\pm$.
In the following $\mathcal{M}^\phi_{\mu\nu}$ denotes the amplitude of $H\to Z + \gamma$ in which the
$W^\pm$ are replaced by their Goldstone bosons $\phi^\pm$. The GBET implies~\cite{GBET}:
\be
\lim_{\tau \to \infty} \,\mathcal{M}_{\mu\nu} (\t ,a)= \lim_{\tau \to \infty} \,
\mathcal{M}^\phi_{\mu\nu}(\t ,a)\label{GBET}\,.
\ee

We calculate  the charged ghost contribution in two different ways:
through direct calculations and via the dispersion integral.
Both calculations lead to the same result.\\

\noindent
$\bullet$ There are 3 vertex graphs and 2 selfenergy graphs, shown
in Figs. 3 and 4, that possibly can contribute. We denote the
contribution from the vertex diagrams by ${\cal
M}_{1+2+3,\m\n}^{\phi}$. Following the Feynmann rules for the
$\phi^\pm$-vertices, given in Appendix \ref{secAA}, with direct
calculations using DimReg we learn that the selfenergy graphs do not
contribute, the result is finite and gauge invariant, as expected:
\begin{align}
&\lim_{\tau \to \infty} \,{\cal M}_{\m\n}^{\phi} = \lim_{\tau \to \infty} \,{\mathcal M}_{1+2+3,\m\n}^{\phi} = \,\lim_{\t\to \infty}\,\frac{eg^2 \cos\theta_W}{8\pi M}\,\frac{1}{2\pi}\,\frac{4a\tau-2\tau}{\t -a}\,{\cal P}_{\m\n} =  -\,\frac{eg^2 \cos\theta_W}{8\pi M}\,\frac{1}{2\pi}\,2(1-2a)\,{\cal P}_{\m\n}\label{GB1} .
\end{align}
Following the GBET, Eq. (\ref{GBET}),
 equations (\ref{Ffull}), (\ref{Finfty}) and (\ref{GB1}) determine the constant ${\cal C}(a)$:
\be
 {\cal C}(a)=2(1-2a) .\label{C}
  \ee
  The details of the calculations are
given in  Appendix~\ref{Higgs_ghost_contr}.

\noindent Thus, our result for the invariant amplitude $\cal F$,
Eqs.~(\ref{Ffull}), (\ref{amplitude_F}) and (\ref{C}),
 completely coincides with the result for the same amplitude ${\cal F}_{\text{DimReg}}$
 obtained in $R_\xi$-gauge with DimReg, Eq.~(\ref{differs}).\\

\noindent
$\bullet$ However, as the goal of our approach with the dispersion integrals is to obtain the amplitude using only finite quantities,
we shall obtain the Goldstone-boson contribution by using the dispersion method.

Analogously to Eq.(\ref{ReM}), we single out the coupling constants
(see the Feynman rules in Appendix \ref{secAA}) and define the
invariant part ${\cal F}^\phi$ of  the  decay amplitude ${\cal
M}_{\m\n}^{\phi}$  in the Higgs-Goldstone boson scalar theory:
\begin{align}
\label{M_phi} {\cal M}_{\m\n}^{\phi}(\t ,a)
= -\frac{eg^2\cos\theta_W}{8\pi M}\, \frac{M_H^2}{4M^2}\,{\cal
F}^\phi (\t ,a) \,{\cal P}_{\m\n}.
\end{align}

We shall apply the dispersive approach (without subtraction) to
the function ${\cal F}^\phi (\tau,a)$. In order to
obtain the  form factor $\tau {\cal F}^\phi (\tau ,a)$ that
enters the amplitude ${\cal M}^\phi_{\m\n}$, Eq. (\ref{M_phi}), we
must multiply the result for  ${\cal F}^\phi (\tau ,a)$ by $\tau$.
 (The same strategy was elaborated
for the $H\rightarrow \gamma + \gamma$ process in \cite{MV}.)

In general,  a constant term can, of course, be always added and in
order to fix the subtraction constant  some additional physical
boundary conditions are required. In contrast to the SM, where the
GBET is a boundary condition that fixes the subtraction constant, in
the Higgs-Goldstone scalar theory there are no asymptotic theorems
one could refer to.

However, the GBET allows to define a boundary condition
for ${\cal F}^\phi (\tau,a)$, as well.  According to the
GBET, the constant ${\cal C}(a)$ is obtained as the large-$\tau$
limit, Eq. (\ref{GBET}), which in terms of the form factors reads:
\begin{align}
\lim_{\tau \rightarrow \infty} 2\pi {\cal F} (\tau,a) = \lim_{\tau
\rightarrow \infty} 2\pi\, [\tau {\cal F}^\phi (\tau,a)] = {\cal C}(a).
\end{align}
Since  ${\cal C}(a)$ is a finite quantity, the structure of Eq.
(\ref{M_phi}) and more precisely  the presence of the factor
$M_H^2$ in the coupling, implies that the large-$\tau$ behavior of
the function $ {\cal F}^\phi(\t ,a)$ is of the form $ {\cal F}^\phi(\t ,a) \sim {\cal O}
(\tau^{-x})$, with $x \geq 1 $. Therefore, the value of the integral
$(1/\pi)\int_{ARC}\text{d}y {\cal F}^\phi (y ,a)/(y-\tau)$ over the
infinite arc in the complex $\tau$-plane, is zero. This, and the fact
that the dispersion integral (see Eq.~(\ref{Fphifull}) bellow) is  convergent, implies that the
dispersion relation applied for ${\cal F}^\phi (\tau ,a)$ does not
need a subtraction.

The absorptive part ${\cal A}^\phi (\t ,a)$ of the  function ${\cal F}^\phi (\t ,a)$
is obtained via the Cutkosky rules from the cut diagrams in
Fig.~\ref{Feynman_Diagrams_phi}.
 Evidently the selfenergy graphs, see  Fig.~\ref{Feynman_Diagram_No_Contribution_phi},
 have no absorptive parts. We obtain (see Appendix~\ref{Higgs_ghost_contr}):
\begin{align}
\Im m\,{\cal M}_{\m\n}^{\phi}(\t ,a)=-\frac{eg^2\cos\theta_W}{8\pi
M}\, \frac{M_H^2}{4M^2} \,{\cal A}^\phi (\t ,a) \,{\cal P}_{\m\n},
\quad \text{with} \quad
{\cal A}^\phi (\t ,a)=(1-2a)\,
\frac{2\a\b -\ln\frac{1+\b}{1-\b}}{2\,(\t -a)^2}. \label{ImMphi}
\end{align}

The expression for  the function ${\cal F}^\phi  (\t ,a) $, valid in the whole $\t$-interval, is obtained via the dispersion integral:
\begin{align}
& {\cal F}^\phi (\t ,a) = \frac{1}{\pi}\,\int \frac{{\cal A}^\phi (y,a)}{y-\t } \text{d}y =  \frac{1-2a}{2\pi}\,\left(4aI_2(\t ,a)-2J_2(\t ,a)\right)\, .\label{Fphifull}
\end{align}
where $I_2(\t ,a)$ and $J_2(\t ,a)$ are convergent and given in
Appendix C.

 In the limit $\t \to \infty$ ($M\to 0$) we have
$I_2(\t ,a)\to 1/(2a (\t -a))$ and $J_2(\t ,a) \to \infty$, and we obtain:
\be
\lim_{\t\to \infty}{\cal F}^\phi (\t ,a)=  \frac{2\,(1-2a)}{2\,\pi\,(\t -a)} .
\ee
Thus, our final result in the limit $M \to 0$ ($\t\to\infty$) is:
\begin{align}
\lim_{\t \to \infty}{\cal M}_{\m\n}^\phi &= \,\lim_{\t \to
\infty}\frac{eg^2 \cos\theta_W}{8\pi M}\,
\,\frac{4a\tau-2\tau}{2\pi\,(\t -a)}\,{\cal P}_{\m\n}=  -\,\frac{eg^2 \cos\theta_W}{8\pi M}\,\,\frac{2(1-2a)}{2\pi}\,{\cal P}_{\m\n} \, ,
\end{align}
which completely coincides with Eq. (\ref{GB1}).

\section{The decay width of $H\to Z + \gamma \,$}

A good approximation for the total width of the Higgs decay into $Z+\gamma$
is given by the contributions of the $W$-boson and the top-quark loops (cf. \cite{BH}):
\begin{align}
&\Gamma(H\rightarrow \, Z + \gamma) = \frac{M_H^3}{32\pi} \,
\left(1 - \frac{M_Z^2}{M_H^2}\right)^3 \,  \left[\frac{eg^2}{(4\pi)^2M}\right]^2 \left| - \cos\theta_W[2\pi\mathcal{F}_W(\tau)] +
\frac{2 \left(3-8\sin^2\theta_W\right)}{3\cos\theta_W} \, [2\pi\mathcal{F}_t(\tau_t)]\right|^2,
\end{align}
where $\mathcal{F}_t(\tau_t)$ stands for the sum of the $t$-quark one-loop diagrams:
\begin{align}
& 2\pi\mathcal{F}_t(\tau_t) = \frac{1}{2(\tau_t-a_t)} \left[1 - (1-\tau_t+a_t) F(\tau_t,a_t) +2a_t G(\tau_t,a_t)\right], \\
& \tau_t = \frac{M_H^2}{4m_t^2}, \qquad a_t = \frac{M_Z^2}{4m_t^2},
\end{align}
and $\mathcal{F}_W(\tau)$ stands for the sum of the $W$-boson one-loop diagrams.\\

Further, we identify $\mathcal{F}_W(\tau)$ with the  amplitude obtained with the dispersion integral,
Eq. (\ref{Ffull}), in which the unsubtracted part is given in (\ref{amplitude_F})
 and ${\cal C}(a)$ in (\ref{C}):
$\mathcal{F}_W(\tau) =\mathcal{F}(\tau,a)$.
This implies that at the measured value for the Higgs mass $ M_H=125.09\,
\text{GeV}$, using  $m_t=172.44\, \text{GeV}$ for the mass of the top-quark, we obtain the following value for the expected decay width:
\begin{align}
 \Gamma(H \rightarrow \, Z + \gamma) = 8.1 \,\text{KeV} \, .
\end{align}

  If,  however, $\mathcal{F}_W(\tau)$ was identified to the unsubtracted
amplitude $\mathcal{F}_W(\tau)= \mathcal{F}_{\text{un}}(\tau,a)$,
 Eq.~(\ref{amplitude_F}), the value for the decay width of $H\to Z + \gamma$ would be  about 20\,\%
 smaller which, as we showed, seems not to be the correct result.




\section{Concluding remarks}

We have calculated the  $W$-boson induced corrections to the decay  $H\to Z + \gamma$ in the Standard Model
in the unitary gauge using the  dispersion-relation approach. This approach is very attractive as it deals only
with finite quantities and thus does not involve any uncertainties related to  regularization.
However, the problem
with the dispersion method is that it determines the amplitude merely up to an additive subtraction constant.

In accordance with this arbitrariness, we calculate the amplitude in two approaches: 1) without subtraction and 2) with subtraction.
We use the the zero-mass limit at $M_W\to 0$ as determined by the GBET, to determine the subtraction constant.
In this latter case   we perform the calculations in two ways: i) through direct calculations of the  amplitude determined by the GBET,
using DimReg, and ii) via the dispersion method,  starting from the absorptive part of the  amplitude, and thus using only finite quantities.
The two completely different calculations lead us to the same subtraction constant!

Furthermore, we also calculated the amplitude  in the commonly used $R_\xi$-gauge class using
dimensional regularization as regularization scheme and compared the result to the one obtained via the dispersion method.
The $R_\xi$-gauge result completely coincides with the dispersion method together with the subtraction term determined by the GBET.

Thus, we have shown that the dispersion-relation approach, with a subtraction determined by the GBET,
 presents an alternative method for calculating the $H\to Z + \g$ amplitude (and for $H\to \g + \g$ as also shown in \cite{HS})
 to the commonly used $R_\xi$-gauge technique. However,  the dispersion method  has two important advantages: 1) it deals only with finite quantities and thus is free of uncertainties related to the choice of regularization and 2) it's much simpler -- working in the unitary gauge effectively we deal with only 3 Feynman diagrams, while in the $R_\xi$-gauge one has to consider more than 20 graphs.\\

\acknowledgments
This paper was initiated in a conversation of I.~Todorov and T.~T.~Wu, though we do not share their point of view. We are thankful to I. Todorov for his interest and useful discussions.
E.~Ch. is thankful for the hospitality of HEPHY, Vienna and
acknowledges the support by Grant 08-17/2016  of the Bulgarian Science
Foundation.

\begin{appendix}

 \section{Feynman rules} \label{secAA}

The Feynman rules for building ${\cal M}_{\mu\nu}$, Eq. (\ref{MmunuW}), are presented in the subsections \ref{secAA_1}, \ref{secAA_2}.
The Feynman rules needed for the calculation of the constant ${\cal C}(a)$, defined by (\ref{Ffull}), are presented
in the subsections \ref{secAA_2}, \ref{secAA_3}.\\

\noindent
In all vertex Feynman diagrams it is assumed that all momenta flow into the vertex.

\subsection{Feynman rules involving $W$-boson in the unitary gauge} \label{secAA_1}

\begin{tabular}{l r}
   \quad \, \includegraphics[width=0.25\textwidth]{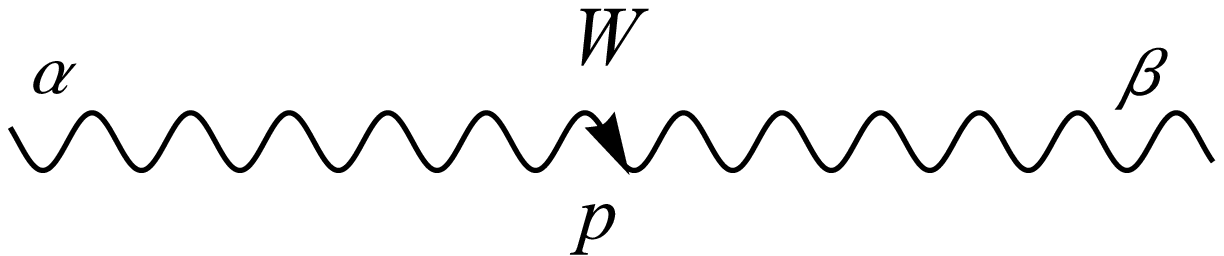} &
    \begin{tabular} {c}
    $ \qquad\qquad\qquad \quad \displaystyle
    \frac{\i}{p^2 - M^2+\i\epsilon} \left(-g^{\alpha\beta} + \frac{p^{\alpha}p^{\beta}}{M^2}\right)
    $ \\
    \, \\ \,
\end{tabular}
\end{tabular}

\begin{tabular}{l r}
    \includegraphics[width=0.25\textwidth]{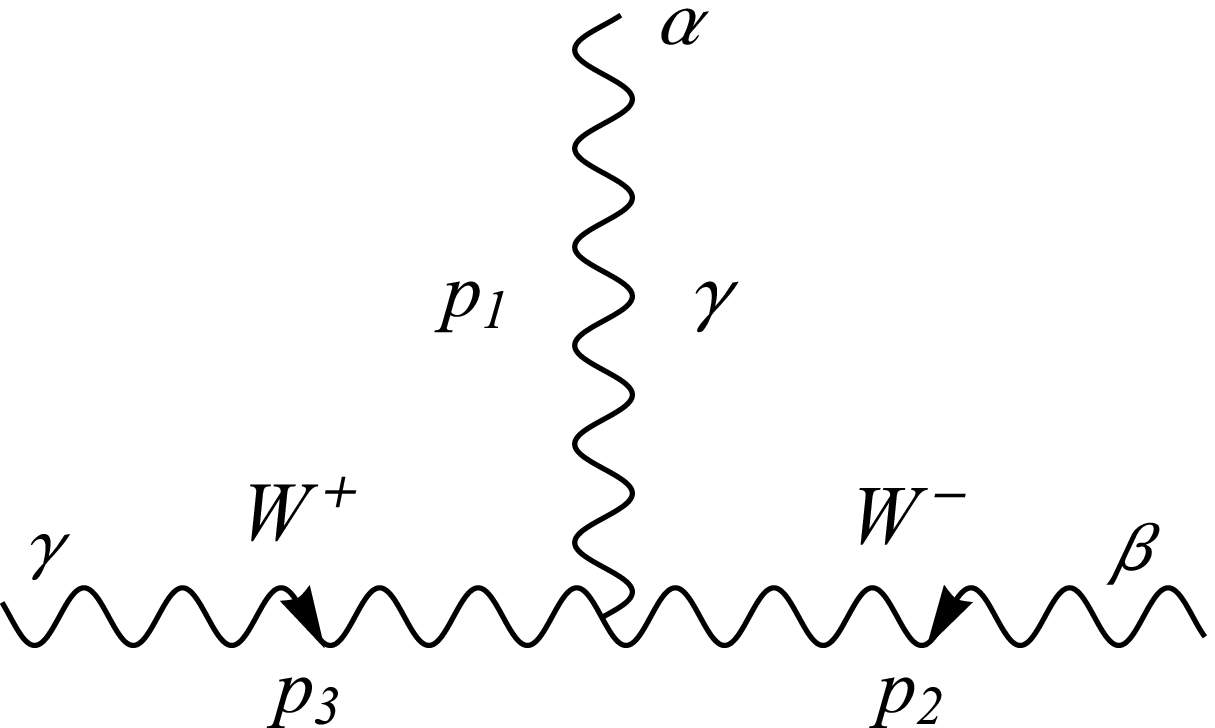} &
    \begin{tabular}{c}
    $ \displaystyle \i eV_{\alpha\beta\gamma}(p_1, p_2, p_3) = \qquad\qquad\qquad\qquad $ \\
    $ \displaystyle \qquad\qquad \, \quad \i e\left[(p_2 - p_3)_{\alpha}g_{\beta\gamma} +
    (p_3 - p_1)_{\beta} g_{\gamma\alpha} + (p_1 - p_2)_{\gamma}g_{\alpha\beta} \right] $ \\
     \, \\ \, \\ \,
    \end{tabular}  \\
\end{tabular}

\begin{tabular}{l r}
    \includegraphics[width=0.25\textwidth]{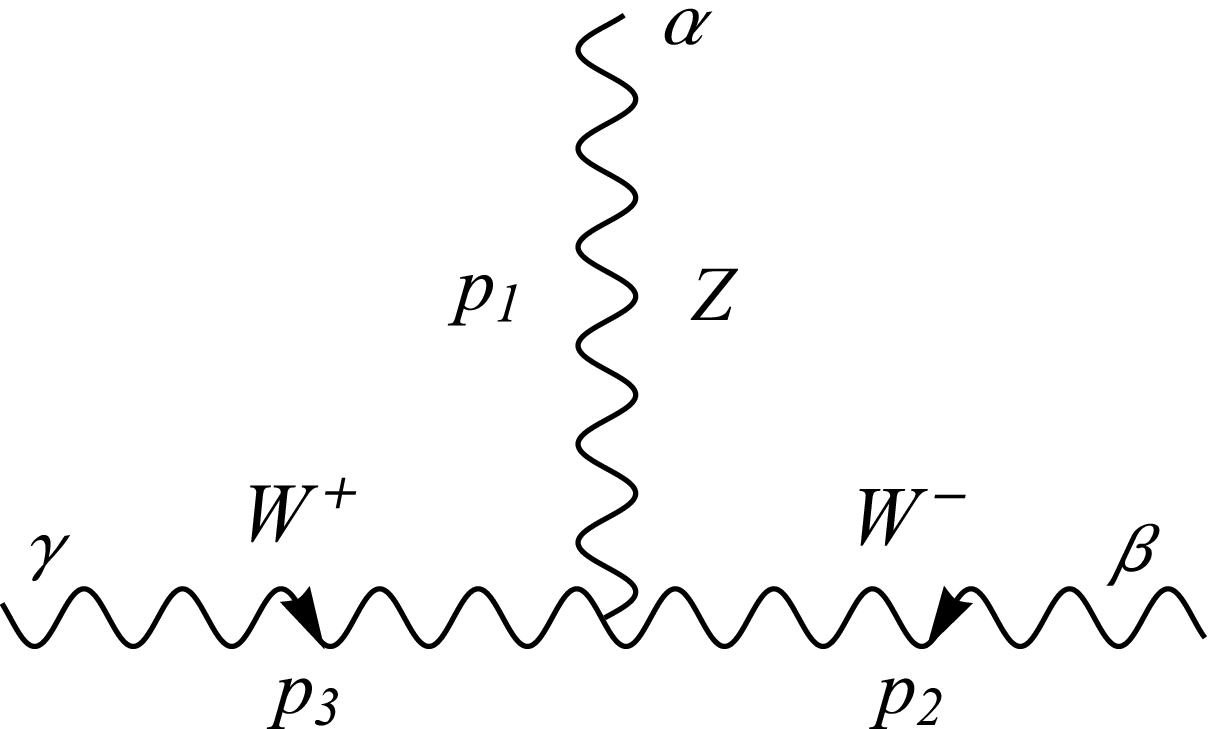} &
    \begin{tabular} {c}
    $ \displaystyle \i g\cos\theta_{W} V_{\alpha\beta\gamma}(p_1, p_2, p_3) =
    \qquad\qquad\qquad\qquad $\\
    $ \displaystyle \qquad\qquad\quad \i g\cos\theta_{W}\left[(p_2 - p_3)_{\alpha}g_{\beta\gamma} +
    (p_3 - p_1)_{\beta}g_{\gamma\alpha} + (p_1 - p_2)_{\gamma}g_{\alpha\beta} \right] $ \\
     \, \\ \, \\ \,
    \end{tabular}  \\
\end{tabular}

\begin{tabular}{l r}
    \includegraphics[width=0.23\textwidth]{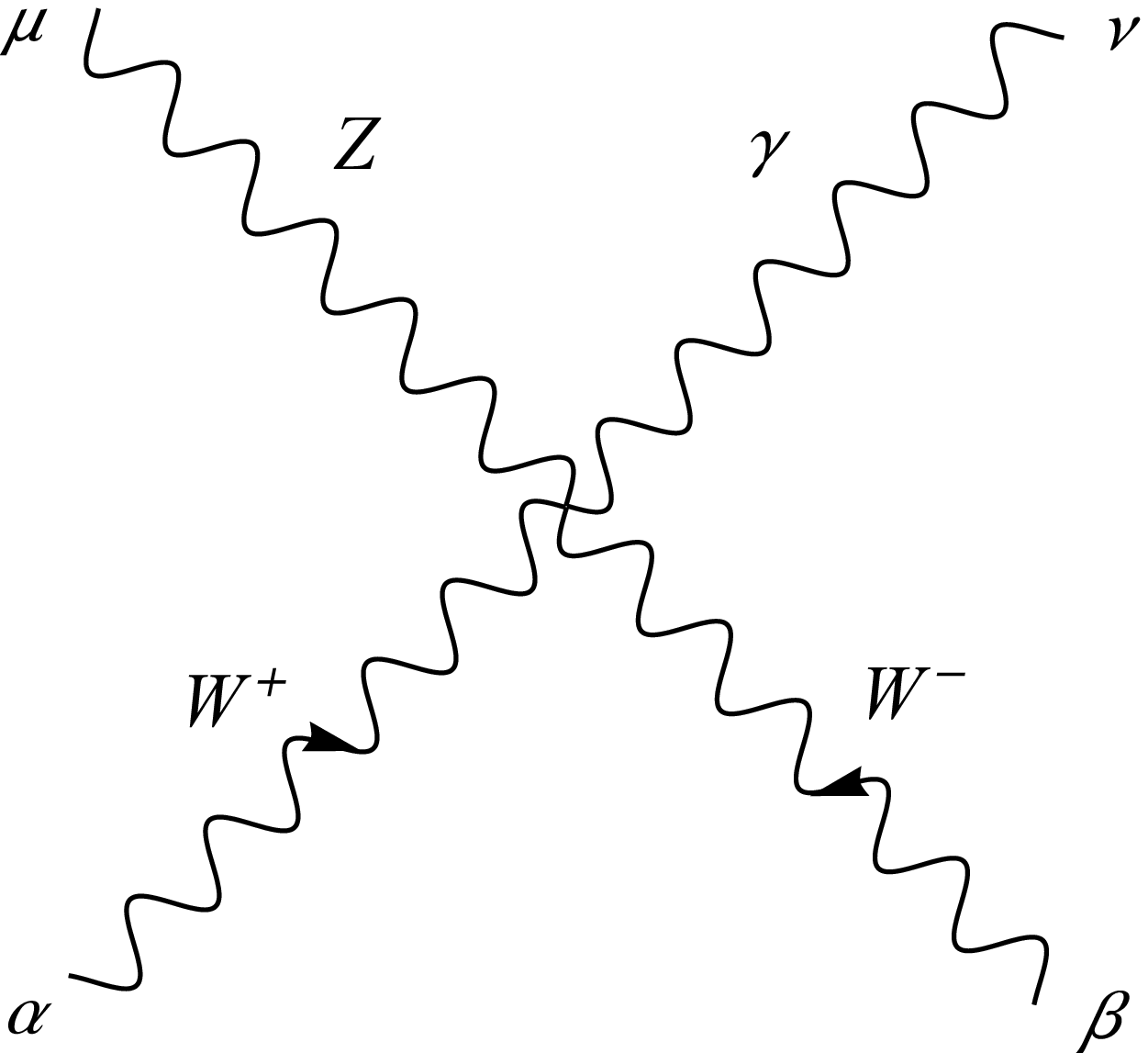} &
    \begin{tabular}{c}
    $  \displaystyle
    -\i eg\cos\theta_{W} V_{\alpha\beta\mu\nu} =  \qquad\qquad$ \\
    $ \qquad \quad \quad \quad \displaystyle -\i eg\cos\theta_{W}\left[2g_{\alpha\beta}g_{\mu\nu}
     - g_{\alpha\mu}g_{\beta\nu} - g_{\alpha\nu}g_{\beta\mu}\right] $ \\
    \, \\ \, \\ \,
    \end{tabular}  \\
\end{tabular}

\begin{tabular}{l r}
    \includegraphics[width=0.25\textwidth]{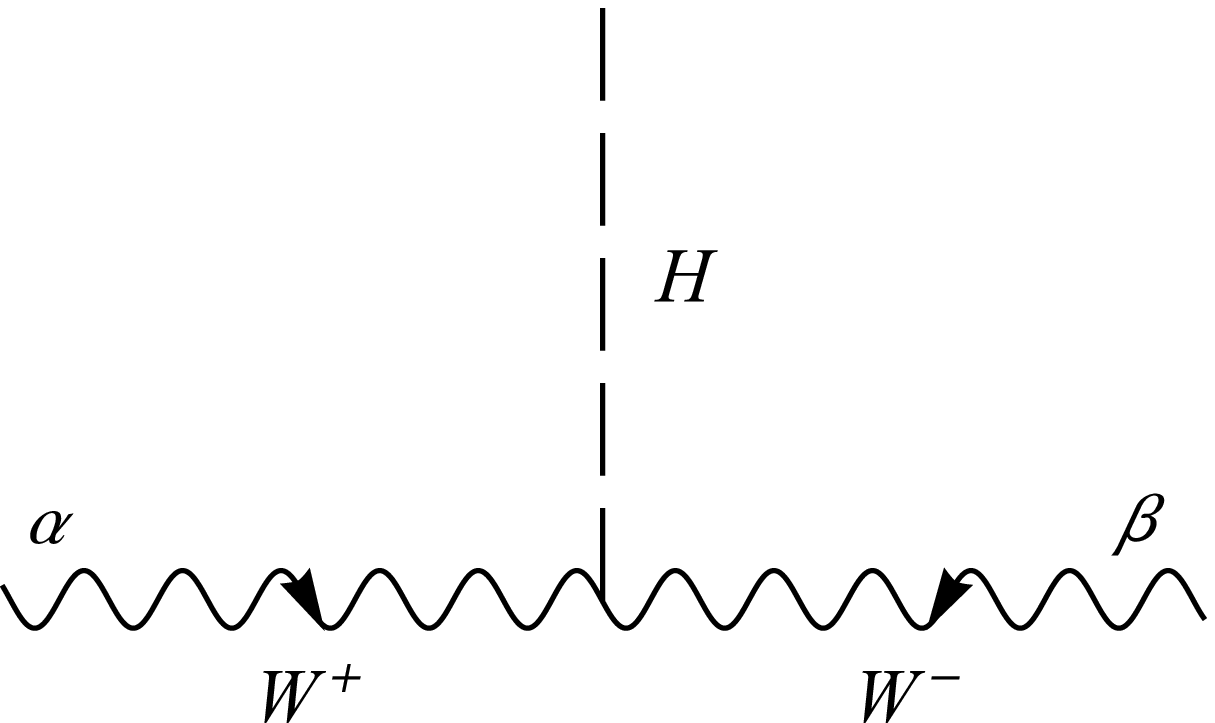} &
    \begin{tabular}{c}
    $ \qquad\qquad\qquad\qquad\qquad\qquad\qquad \displaystyle
    \i gMg_{\alpha\beta} $ \\
    \, \\ \, \\ \,
    \end{tabular}  \\
\end{tabular}

\subsection{Feynman vertex rule for the triple Higgs -- \boldmath $Z$-boson interaction} \label{secAA_2}

\begin{tabular}{l r}
    \includegraphics[width=0.25\textwidth]{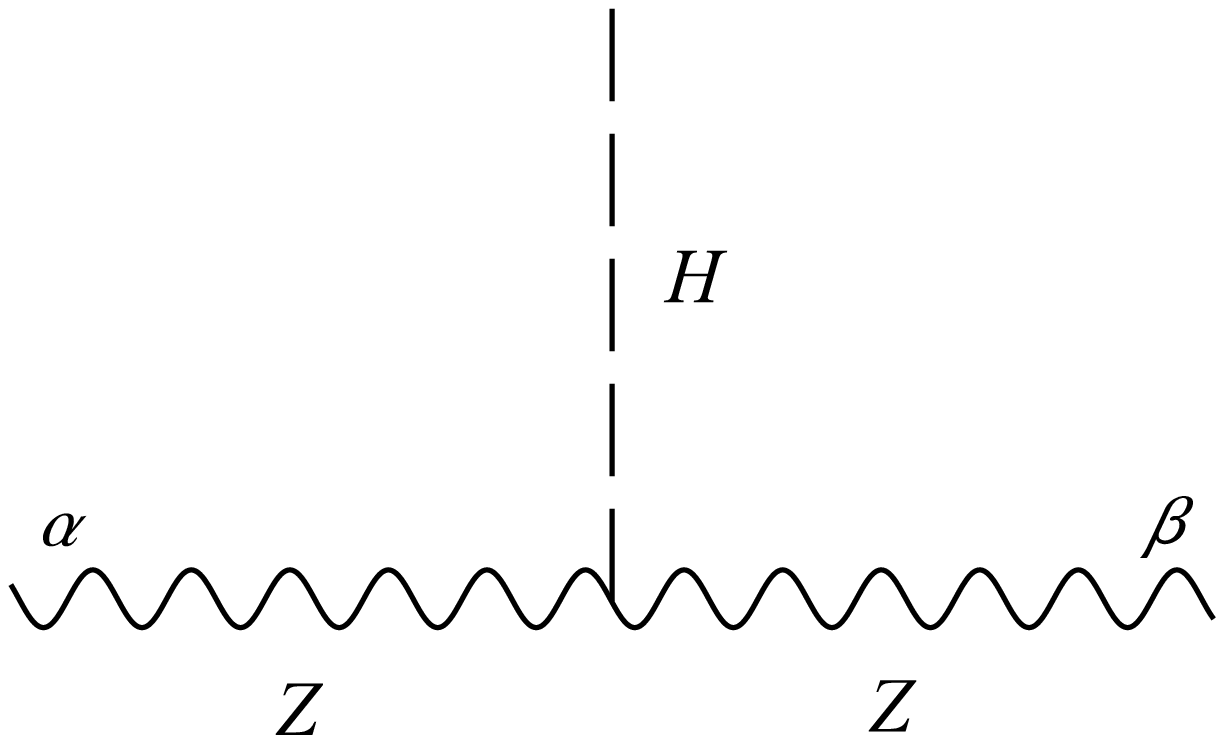} &
    \begin{tabular}{c}
    $ \qquad\qquad\qquad\qquad\qquad\qquad\qquad \displaystyle
    \i \, \frac{g}{\cos\theta_W} \, M_Zg_{\alpha\beta} $ \\
    \, \\ \, \\ \,
    \end{tabular}  \\
\end{tabular}

\subsection{Feynman rules involving the charged Higgs ghost in the  \boldmath $R_{\xi}$ gauge} \label{secAA_3}

\begin{tabular}{l r}
   \quad \, \includegraphics[width=0.25\textwidth]{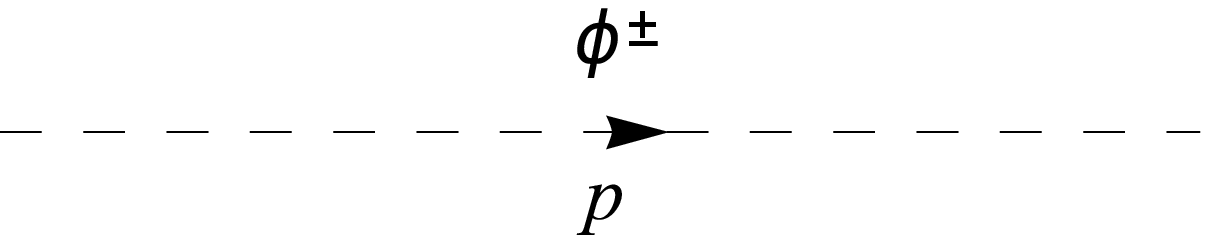} &
    \begin{tabular} {c}
    $ \qquad\qquad\qquad\qquad \quad \displaystyle
    \frac{\i}{p^2 - \xi_W M^2+\i\epsilon}
    $ \\
    \, \\ \,
\end{tabular}
\end{tabular}

\begin{tabular}{l r}
    \includegraphics[width=0.25\textwidth]{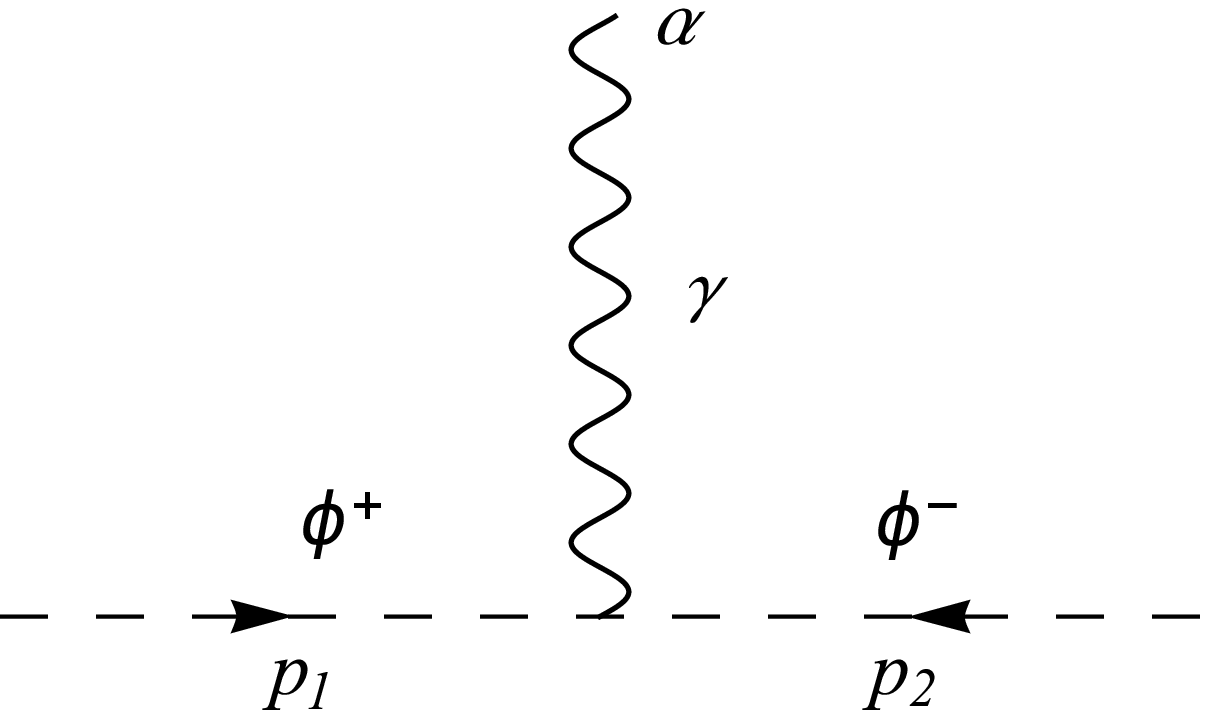} &
    \begin{tabular}{c}
    $ \displaystyle\qquad\qquad\qquad\qquad \qquad\qquad \i e (p_1-p_2)_{\alpha}  $ \\
        \, \\ \,
    \end{tabular}  \\
\end{tabular}

\begin{tabular}{l r}
    \includegraphics[width=0.25\textwidth]{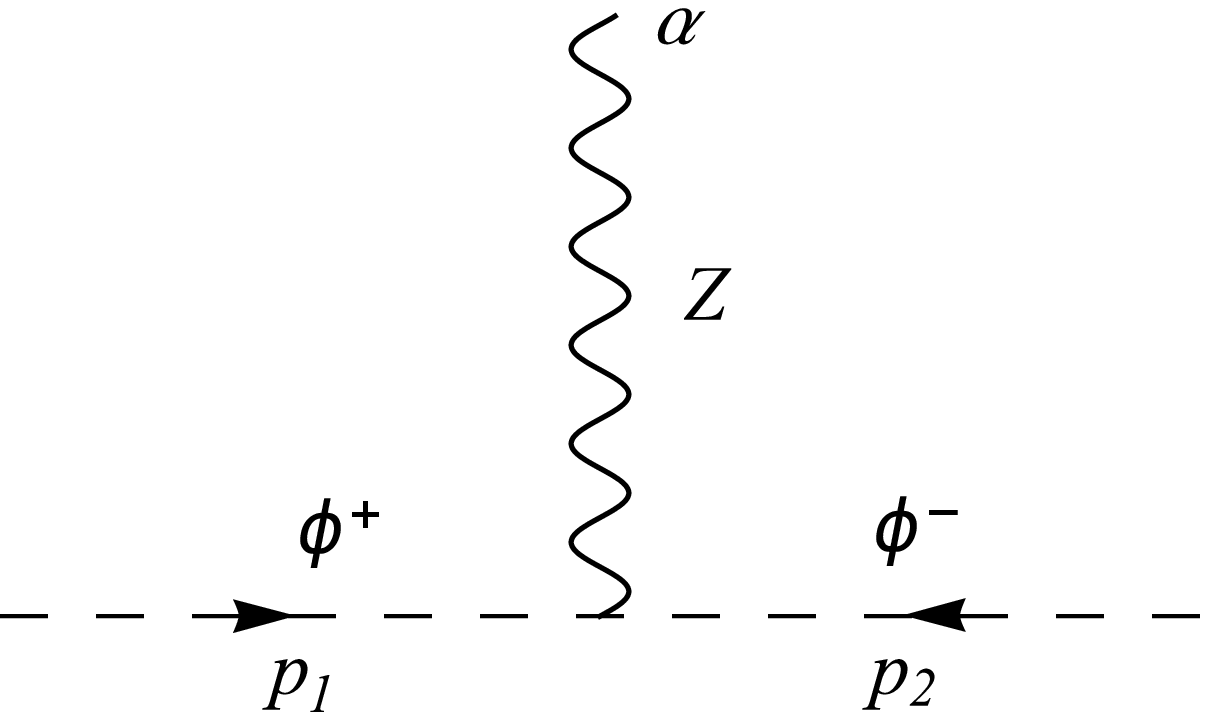} &
    \begin{tabular} {c}
    $ \displaystyle\qquad\qquad\qquad  -\i \,g\, x_W\, (p_1-p_2)_{\alpha}, \qquad
    x_W = -\,\frac{\cos 2\theta_W}{2\cos\theta_W} $ \\
           \, \\ \,
    \end{tabular}  \\
\end{tabular}

\begin{tabular}{l r}
    \includegraphics[width=0.23\textwidth]{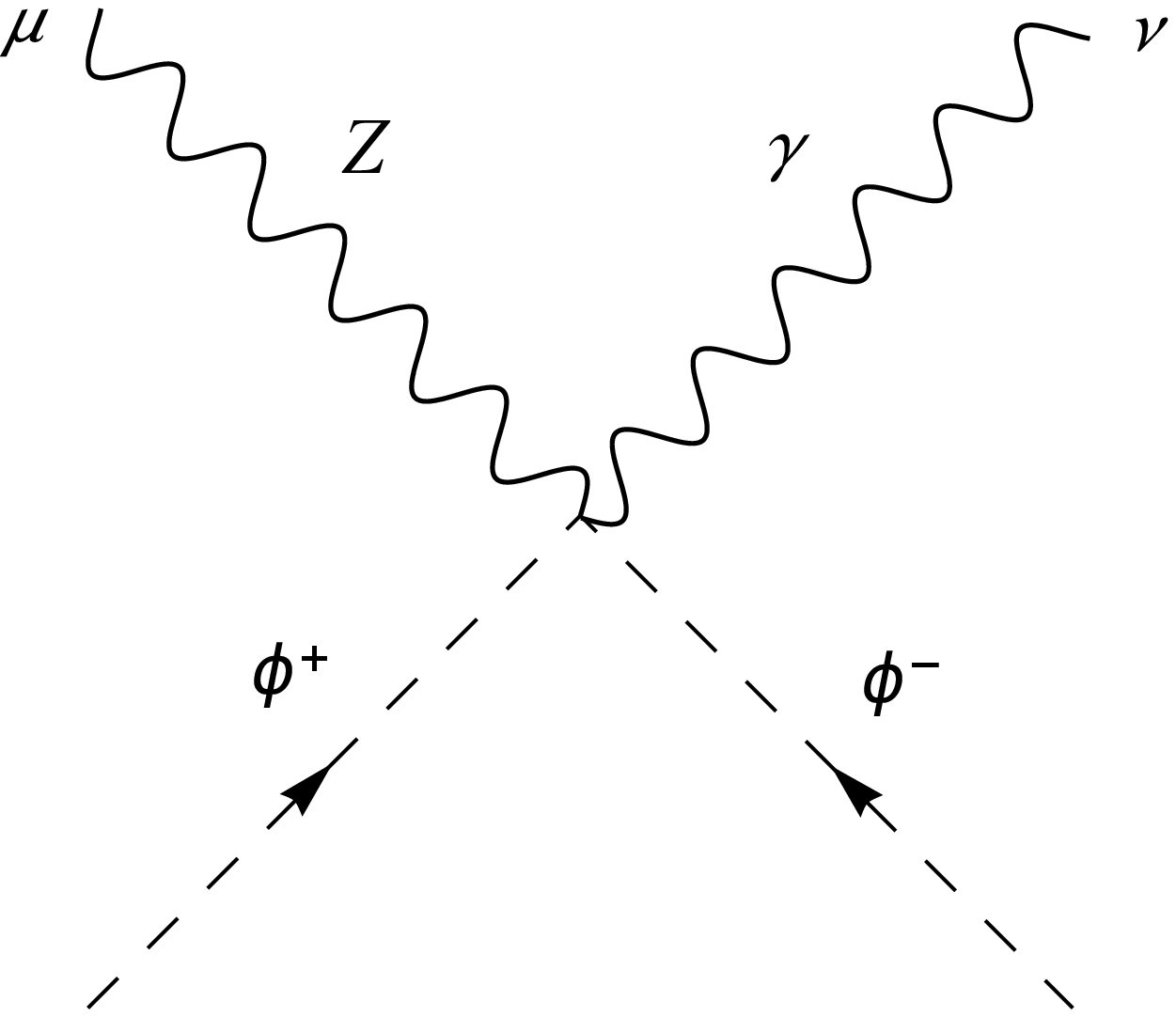} &
    \begin{tabular}{c}
    $  \displaystyle \qquad\qquad\qquad\qquad\qquad\qquad\qquad
     -2\,\, \i\, e\,g\,x_W\, g_{\mu\nu} $ \\
    \, \\ \, \\ \,
    \end{tabular}  \\
\end{tabular}

\begin{tabular}{l r}
    \includegraphics[width=0.25\textwidth]{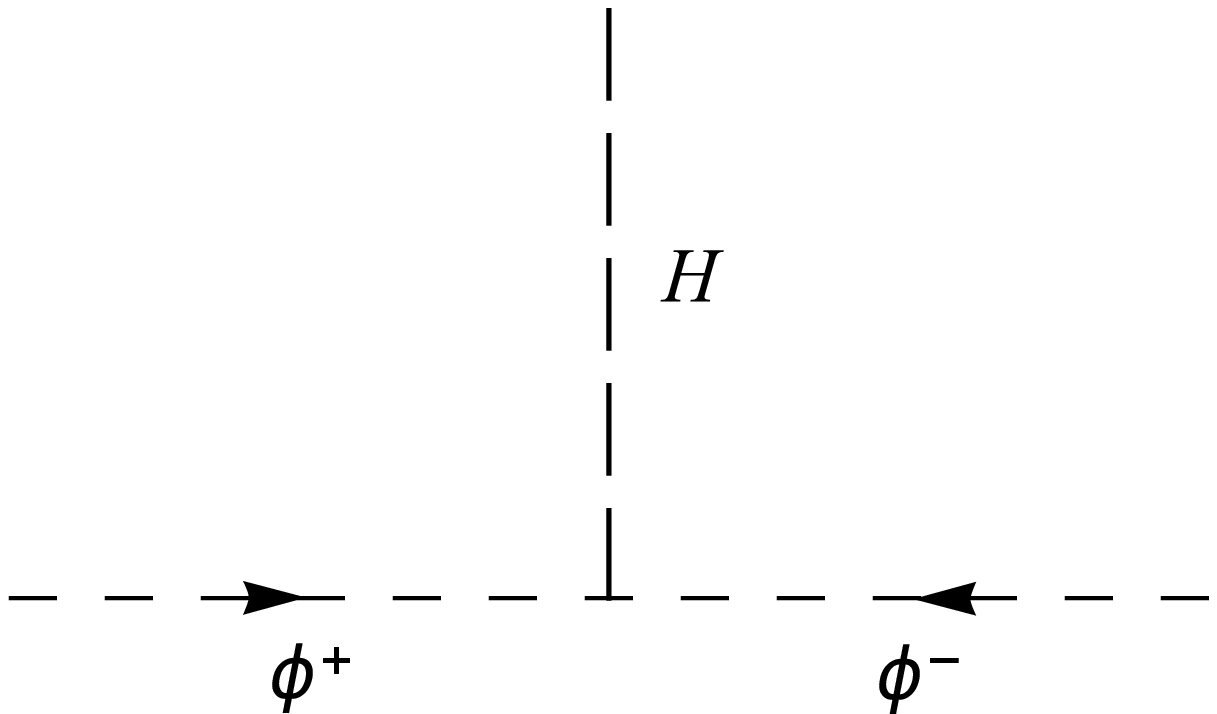} &
    \begin{tabular}{c}
    $ \qquad\qquad\qquad\qquad\qquad\qquad\qquad \displaystyle
    -\i\, g \, \frac{M_{H}^2}{2M} $ \\
    \, \\ \, \\ \,
    \end{tabular}  \\
\end{tabular}

\section{Integrals for the absorptive part  \boldmath $\mathcal{A}(\tau)$}\label{secAB}

\noindent
Here we give the integrals involved in computation of the absorptive part
$\mathcal{A}(\tau)$ of the amplitude.\\

\noindent
The calculations are done in the rest frame of the Higgs boson, with $z$-axis
taken along $\text{k}_1$, the kinematics as given in Sec. \ref{AbsorptivePartAmplitude}. We have used also
the following relations:
\begin{align}
k^2 = - \vert\text{k}\vert^2 = - M^2(\tau-1), \, (k\cdot p) = 0, \,
(k\cdot v) = -2M^2(\tau-a) \, \beta\cos\theta.
\end{align}
The evaluation of $\mathcal{A}(\tau)$ is reduced to the following integrals:
\begin{align}
1.\qquad\Im m \int \frac{\text{d}^4k}{(2\pi)^4} \, \frac{\i}{D_1D_3} = -\frac{\beta}{8\pi},
\end{align}
\begin{align}
2.\qquad & \Im m \int \frac{\text{d}^4k}{(2\pi)^4} \, \frac{\i}{D_1D_2D_3} = \frac{\beta}{32\pi M^2 (\tau-a)} \, I , \\
\nn
& I = \int_{-1}^{+1}\frac{\text{d}x}{1 - \beta x} = \frac{1}{\beta}\ln\left(
\frac{1 + \beta}{1 - \beta}\right) ,
\end{align}

\begin{align}
3.\qquad & \Im m \int \frac{\text{d}^4k}{(2\pi)^4} \, \frac{\i k_{\mu}}{D_1D_2D_3} =
- \frac{\beta^2\tau}{64\pi M^2 (\tau-a)^2} \,
\left(\frac{a}{\tau} \, p_{\mu} - v_{\mu}\right) \,  J , \\
\nn
& J = \int_{-1}^{+1}\frac{x\text{d}x}{1 - \beta x} = \frac{1}{\beta} \left(I-2\right) ,
\end{align}

\begin{align}
4.\qquad \Im m \int \frac{\text{d}^4k}{(2\pi)^4} \, \frac{\i k_{\mu}k_{\nu}}{D_1D_2D_3}
=  L_1\left(g_{\mu\nu}-\frac{p_\mu p_\nu }{p^2}\right) + L_2\left(\frac{a}{\tau}
\, p_{\mu} - v_{\mu}\right) \left(\frac{a}{\tau} \, p_{\nu} - v_{\nu}\right)
\end{align}
where
\begin{align}
& L_1 = - \frac{ \beta^3 \tau}{64\pi (\tau-a)} \, \left( I - K \right) , \\
& L_2 = - \frac{ \beta^3 \tau^2 }{256\pi M^2 (\tau-a)^3} \, \left( I - 3K \right) , \\
& K = \int_{-1}^{+1}\frac{x^2\text{d}x}{1 - \beta x} = \frac{1}{\beta^2} \left(I-2\right) .
\end{align}
We recall the notation:
\begin{align}
D_1=\left(k+\frac{p}{2}\right)^2-M^2,\quad D_2=\left(k-\frac{v}{2}\right)^2-M^2,\quad
D_3=\left(k-\frac{p}{2}\right)^2-M^2.
\end{align}

\section{Integrals for the real part of  \boldmath $\mathcal{F}(\tau)$}\label{secAC}

\noindent
The invariant amplitude $\mathcal{F}(\tau)$, Eq. (\ref{amplitude_F}),  is
 a linear combination of the  dispersion integrals $I_i$ and $J_i$:
\begin{align}\label{FI2_9}
2\pi\mathcal{F}(\tau)  =& \,\,4a\,I_1(\tau, a) + 6a\,I_2(\tau, a) - 8a^2\, I_3(\tau, a) + 2(3-2a)\,J_1(\tau, a) \notag \\
& + 4a^2\, J_2(\tau, a)  - 3\,J_3(\tau, a) - 3a\, J_4(\tau, a) ,
\end{align}
which we list below. We distinguish  two types of integrals:\\
\begin{enumerate}
\item
Integrals with $\beta$:
\be
\beta\equiv \sqrt{1-y^{-1}}.
\ee
They are expressed in terms of the integral $I_0(x)$, or equivalently of the elementary function $g(x)$:
\begin{align}
I_0(\tau) = \frac{1}{2} \int_{1}^{\infty} \frac{\beta}{(y-\tau)y} \, \text{d}y =
\frac{1}{\tau} [1 - g(\tau)],
\end{align}
where
\begin{align}
g(\tau )\equiv \sqrt{\tau^{-1}-1} \arcsin\sqrt{\tau }.
\end{align}

For the other integrals we have:
\begin{align}
I_1(\tau,a) &= \, \frac{1}{2} \int_1^{\infty} \frac{\beta}{(y-\tau)(y-a)} \, \text{d}y =  \frac{g(a)-g(\tau)}{\tau-a} , 
\end{align}

\begin{align}
I_2(\tau,a) &=
\frac{1}{2} \int_1^{\infty} \frac{\beta}{(y-\tau)(y-a)^2} \, \text{d}y = \frac{\partial}{\partial a} I_1(\tau,a) =\notag \\
& = \frac{1}{\tau-a} \left\{ I_1(\tau,a) - \frac{1}{2(1-a)}
\left[1-I_0(a)\right] \right\},
\end{align}

\begin{align}
I_3(\tau,a) &= \frac{1}{2} \int_1^{\infty} \frac{\beta y}{(y-\tau)(y-a)^2} \,\text{d}y =   I_1(\tau,a) + aI_2(\tau,a);
\end{align}

\item
Integrals with the  logarithm $l_{\beta}$:
\begin{align}
 l_{\beta}\equiv \ln\frac{1+\beta}{1-\beta}.
\end{align}
They are expressed in terms of the integral $J_0(x)$, or equivalently of the elementary function $f(x)$:
\begin{align}
J_0(\tau) = \frac{1}{2} \int_{1}^{\infty} \frac{l_{\beta}}{(y-\tau)y} \, \text{d}y =
 \frac{f(\tau)}{\tau},
\end{align}
where
\begin{align}
f(x)\equiv \arcsin^2 \sqrt{x}.
\end{align}

We have:
\begin{align}
J_1(\tau,a) &= \frac{1}{2} \int_1^{\infty} \frac{l_{\beta}}{(y-\tau)(y-a)} \, \text{d}y =  \frac{f(\tau)-f(a)}{\tau-a}, 
\end{align}
\begin{align}
J_2(\tau,a) &= \frac{1}{2}
\int_1^{\infty} \frac{l_{\beta}}{(y-\tau)(y-a)^2} \, \text{d}y =\frac{\partial}{\partial a} J_1(\tau,a)  = \frac{1}{\tau-a} \left[ J_1(\tau,a) - \frac{g(a)}{1-a} \right],
\end{align}

\begin{align}
J_3(\tau,a) &= \frac{1}{2} \int_1^{\infty} \frac{l_{\beta}}{(y-\tau)(y-a)y} \, \text{d}y =  \frac{1}{\tau-a} \left[J_0(\tau) - J_0(a)\right],
\end{align}

\begin{align}
J_4(\tau,a) &=  \frac{1}{2}
\int_1^{\infty} \frac{l_{\beta}}{(y-\tau)(y-a)^2y} \, \text{d}y = \frac{\partial}{\partial a} J_3(\tau,a) =  \frac{1}{a}\,\, [J_2(\tau,a) - J_3(\tau,a)]
\end{align}

\end{enumerate}

Below we  give the expansions used to determine the limits at $a \to 0$ and $\tau \to a$.
At $\vert a \vert <1$ we have:
\bea
f(a) &=& a  + \frac{1}{3} a^2 +  \frac{8}{45} a^3 + O(a^4), \\
g(a) &=& 1 - \frac{1}{3} \, a  - \frac{2}{15} \, a^2 -\frac{8}{105}\ a^3 + O(a^{7/2}).
\eea

When $\vert \tau - a\vert < 1$ for the functions $F(a,\tau )$ and $G(a,\tau )$,
that enter the amplitude (\ref{amplitude_F}), we have:
\begin{align}
 F(a,\tau ) &=\, \frac{f(\tau)-f(a)}{\tau-a} =  \frac{1}{1 - a} \, g(a) + \frac{1}{4a(1-a)} \,
 \left[1 - \frac{(1 - 2 a)}{1 - a} \, g(a)\right] \, (\tau - a)  \notag \\
& \quad + \frac{1}{8a^2(1 - a)^2} \, \left[2a - 1 + \frac{(8a^2 - 8a + 3)}{3(1 - a)} \, g(a)\right] \, (\tau - a)^2   + O((\tau - a)^3).
\end{align}
\begin{align}
G(a,\tau ) &= \frac{g(\tau)-g(a)}{\tau-a} = \frac{1}{2a} \left[1 - \frac{1}{1-a} \, g(a)\right]  - \frac{1}{8a^2(1-a)^2} \left[2a^2 - 5a + 3 + (4a-3) \, g(a)\right] (\tau-a)  \notag \\
& + \frac{1}{48(1-a)^3a^3} \left[-8a^3 + 34a^2 - 41a + 15  - 3 (8a^2 - 12a + 5) \, g(a)\right] (\tau-a)^2 + O((\tau-a)^3) .
\end{align}

\section{The charged Higgs ghost contribution to \boldmath $H \to Z  + \gamma$} \label{Higgs_ghost_contr}

In Appendix~\ref{secAA_3} all involved Feynman rules are given in
the $R_\xi$ gauge.\\
Here we calculate the charged Higgs ghosts $\phi^\pm$ contribution, according to the GBET.
We use the unitary gauge -- the diagrams in Figs.1 and 2, in which
the virtual $W$-bosons are replaced by the physical scalars
$\phi^\pm$ -- Figs. 3 and 4. Their propagators are that of a  scalar,
with  mass of the $W$-boson, ${\i \over p^2 - M^2 + \i \epsilon}$.
All couplings in Appendix~\ref{secAA_3} are $\xi$-independent and therefore we can take them directly.\\

$\bullet$ First we calculate the constant part of this contribution
evaluating the intergrals by using Feynman parametrization. The
occurring UV~divergent integrals are regularized with dimensional
regularization.\\

\noindent
Based on Fig.~\ref{Feynman_Diagrams_phi}
we get the matrix elements
\begin{align}
{\cal M}^\phi_{1\, \mu\nu}  & = \i e g^2 x_W  {M_H^2 \over 2 M} {1 \over (2 \pi)^4} \,\int {\rm d}^4 k \, {(P_1 + P_2)_\mu  (P_2 + P_3)_\nu \over D_1 D_2 D_3} \, \\
{\cal M}^\phi_{3\, \mu\nu}  & = \i e g^2 x_W  {M_H^2 \over 2 M} {1 \over (2 \pi)^4} \,\int {\rm d}^4 k \, { ( \tilde  P_2 +P_3)_\mu (P_1 + \tilde P_2)_\nu \over D_1 \tilde D_2 D_3} \, \\
{\cal M}^\phi_{2\, \mu\nu}  & = - \i e g^2 x_W  {M_H^2 \over 2 M} {1 \over (2 \pi)^4} \,\int {\rm d}^4 k \, { 2 g_{\mu\nu} \over D_1 D_3} \, .
\end{align}
Here we use $D_i = P_i^2 - M^2, P_1 = k, P_2 = k - k_1, P_3 = k - k_1 - k_2, \tilde P_2 = k - k_2$.
Furthermore, similar to Eq.~(\ref{}) we can write the sum of the three vertex amplitudes as
$2 {\cal M}^\phi_{1\, \mu\nu} + {\cal M}^\phi_{2\, \mu\nu}$, which is
\begin{equation}
{\cal M}^\phi_{1+2+3\, \mu\nu} =   \i e g^2 x_W  {M_H^2 \over M} {1 \over (2 \pi)^4} \,\int {\rm d}^4 k \,
{T \over D_1 D_2 D_3},
\label{M123phi}
\end{equation}
with
\begin{align}
T &= (P_1 + P_2)_\mu  (P_2 + P_3)_\nu  - g_{\mu\nu} D_2 =  4 k_\mu k_\nu - 4 k_\mu k_{1\nu} + (2 k.k_1 + M^2 - M_Z^2 - k^2) g_{\mu\nu}\, .
\end{align}
Using the formula for Feynman parametrization,
\begin{align}
&{1 \over  D_1 D_2 D_3}  =  2 \int_0^1 {\rm d}x_1 \int_0^{1 - x_1} {\rm d}x_2 {1 \over  (x_1 D_1 + x_2 D_2 + (1 - x_1 - x_2) D_3)^3}\, ,
\label{feynman-para}
\end{align}
and by the substitution
$k_\mu \to l_\mu + (1 - x_1) k_{1 \mu} + (1 - x_1 -x_2) k_{2 \mu}$ we get
\begin{align}
&{1 \over  D_1 D_2 D_3} =  \int_0^1 {\rm d}x_1   \int_0^{1 - x_1} {\rm d}x_2 {2 \over   (l^2 - \Delta)^3} , \,\,\Delta = M^2 + 2 k_1.k_2 x_1 (x_1 + x_2 - 1) + M_Z^2  x_1(x_1 - 1) \, .
\end{align}
The two necessary integrals over $l$ are
\begin{align}
\int {\rm d}^4 l {1 \over (l^2 - \Delta_i)^3} & =   - \i \pi^2 {1 \over 2 \Delta_i}\, , \\
\int {\rm d}^4 l {l^2 \over (l^2 - \Delta_i)^3} & =   \i \pi^2  \left(\Delta^{UV} -
{1 \over 2}\right), \quad {\rm with}   \quad   \Delta^{UV}  = {1 \over \epsilon} + {\rm const.}
\end{align}
All odd powers of $l$ vanish due to the symmetric integration and thus will be dropped.
Applying $l_\mu l_\nu = {l^2 \over d}$ with the dimension parameter
$d = 4 - 2 \epsilon$ we get
\begin{align}
T & \to \quad  4 k_{1\nu} k_{2\mu} x_1 (x_1 + x_2 - 1) \, +   \left(M^2 - 2 k_1.k_2 x_1  (x_1 + x_2 - 1) - M_Z^2 x_1^2 +
 \epsilon {l^2  \over 2}\right) g_{\mu\nu}\, .
\end{align}
Integrating over $l$ and neglecting terms of the order
$M^2/M_H^2$ and  $M_Z^2/M_H^2$ we obtain:
\begin{align}
 - {\i \over \pi^2}\int {\rm d}^4 k \, {T \over D_1 D_2 D_3} & = {1 \over 2} \,\,g_{\mu\nu} -
\int_0^1 {\rm d}x_1  \int_0^{1 - x_1} {\rm d}x_2  {- 4 k_{1\nu} k_{2\mu} x_1 x_2 + (M^2 + 2 k_1.k_2 x_1 x_2 - M_Z^2 x_1^2) g_{\mu\nu} \over
M^2 - 2 k_1.k_2 x_1 x_2 + M_Z^2  x_1(x_1 - 1)} = \nonumber\\
&=   - {1 \over k_1.k_2} \, {\cal P}_{\mu\nu} +  \ldots\, .
\end{align}
with $ {\cal P}_{\mu\nu}$ given by Eq.~(\ref{Ptransverse}).
By inserting this result into Eq.~(\ref{M123phi}) we receive for the
leading term of the vertex graphs with the charged ghost:
\begin{align}
\label{resM123ghost1}
\hspace{0cm}
& {\cal M}^\phi_{1+2+3\, \mu\nu}  = e g^2 x_W  {M_H^2 \over M} {\pi^2 \over (2 \pi)^4} {P_{\mu\nu} \over k_1.k_2} =   {e \,g^2\, \cos\theta_W \over 8 \pi \, M}{1 \over 2 \pi}  \left({M_Z^2 \over M^2} - 2\right) {M_H^2  \over M_H^2 - M_Z^2}
  P_{\mu\nu} =  {e \,g^2\, \cos\theta_W \over 8 \pi \, M} {1 \over 2 \pi}
{4 a \tau - 2 \tau \over \tau - a} P_{\mu\nu}\, .
\end{align}
We have used $2 k_1.k_2 = M_H^2 - M_Z^2$, and $x_W/\cos\theta_W = {1
\over 2} \left({M_Z^2 \over M^2} - 2\right) = (2 a - 1)$ after
inserting $\tau = {M_H^2 \over 4 M^2}$ and $a = {M_Z^2 \over 4
M^2}$. Explicit calculations show that the sum of the two
self-energy graphs, given by
Fig.~\ref{Feynman_Diagram_No_Contribution_phi}
vanishes in the unitary gauge using dimensional regularization.\\

\noindent
$\bullet$ Now we derive the imaginary part of the amplitude $ {\cal
M}^\phi_{\mu\nu}$ by applying the Cutkosky cuts to
Fig.~\ref{Feynman_Diagrams_phi}. We have
\begin{align}
\Im m \,\mathcal{M}^\phi_{\mu\nu} &= -\frac{\i}{2} \, (2\mathcal{M}_{1\mu\nu}^{\phi\, C} +
\mathcal{M}_{2\mu\nu}^{\phi\, C} )\, .
\end{align}
where "C" denotes the cut diagrams.
Then $\Im m \,\mathcal{M}^\phi_{\mu\nu}$ can be written as
\begin{align}
&\Im m \,\mathcal{M}^\phi_{\mu\nu}  = e g^2 x_W  {M_H^2 \over 2 M}  \Im m\int {{\rm d}^4 k \over (2 \pi)^4}\,
\i \left( {(P_1 + P_2)_\mu  (P_2 + P_3)_\nu \over D_1 D_2 D_3}
- {g_{\mu\nu} \over D_1 D_3} \right)\,,
\end{align}
with the momenta and denominators defined in section~\ref{section_FeynDiags},
with the substitution for $D_1$ and $D_3$ following Eq.~(\ref{Cutkosky-trans}), and
\begin{equation}
(P_1 + P_2)_\mu  (P_2 + P_3)_\nu = 4 k_\mu k_\nu - 2 k_\mu k_{1 \nu} +  2 k_\nu k_{2 \mu} -  k_{1 \nu} k_{2 \mu} \, .
\end{equation}
By using Appendix~\ref{secAB} we get the result
\begin{equation}
\Im m \,\mathcal{M}^\phi_{\mu\nu}  =  e g^2 x_W  {M_H^2 \over 2 M} {2 a \beta +  \tau  (\beta^2 - 1)\ln{1 + \beta \over 1 - \beta} \over 32 \pi M^2 (\tau - a)^2} {\cal P}_{\mu\nu}\, .
\end{equation}
With $x_W = \cos\theta_W (2 a - 1)$ and $\beta^2 - 1 = -1/\tau$ we
get the result
\begin{equation}
\Im m \,\mathcal{M}^\phi_{\mu\nu}  =   {e g^2 \cos\theta_W  \over
16 \pi M}  {M_H^2 \over 4 M^2} (2 a - 1) {2 a \beta - \ln{1 + \beta
\over 1 - \beta} \over (\tau - a)^2} {\cal P}_{\mu\nu}\, .
\end{equation}

\end{appendix}


\end{document}